\documentclass{appolba}
\usepackage{epsfig}

\begin{document}
\title{Higgs Decay to Photons at Two Loops%
\thanks{Presented at conference ``Physics at LHC", Cracow, Poland,
  July 2006}%
}
\author{Frank Fugel
\address{II. Institut f\"ur Theoretische Physik, Universit\"at Hamburg,\\
         Luruper Chaussee 149, 22761 Hamburg, Germany}
}
\maketitle


\begin{abstract}
The calculation of the two-loop corrections to the partial width
of an intermediate-mass Higgs boson decaying into a pair of photons is
reviewed. The main focus lies on the electroweak (EW)
contributions. The sum of the EW corrections ranges from -4\%~to~0\%
for a Higgs mass between 100~GeV and 150~GeV, while the complete
correction at two-loop order amounts to less than $\pm$1.5\% in this
regime.
\end{abstract}


\section{Introduction}
The standard model (SM) predicts the existence of one scalar particle,
the Higgs boson (H). The Higgs boson is the only particle of the SM
which has not been found until now. Electroweak precision data mainly
collected at CERN LEP and SLAC SLC in combination with the direct
top-quark mass measurement at the Tevatron would favour a light Higgs
boson with a mass below about 200~GeV at the 95\% confidence level,
while the direct search at LEP leads to a lower bound of 114~GeV at
the 95\% confidence level~\cite{ewwg}. This mass range is compatible
with the so-called intermediate-mass range, defined by $M_W\le M_H\le
2M_W$, $M_W$ and $M_H$ being the mass of the W boson and the Higgs
boson, respectively. In this mass regime the decay of the Higgs boson
into a pair of photons is an important detection channel at the Large
Hadron Collider (LHC) due to its clear signature, though the branching
fraction does not exceed 0.3\%. Furthermore, this decay channel is
useful in determining the properties of the Higgs boson. At a future
International Linear Collider (ILC) precision measurements would be
possible. In particular, at the ILC the two photon mode could be made
possible, which allows for the production of Higgs bosons via the
fusion of two photons. This way a precise measurement of $\Gamma(H \to
\gamma \gamma)$, with a precision of 2.3\% for
$M_H=120$~GeV~\cite{Djouadi:2005gi}, would be possible. Also the
CP-properties of the Higgs boson could be studied at the ILC operating
in the two photon mode. A comprehensive review of SM Higgs boson
physics is given in~\cite{Djouadi:2005gi}. The Higgs-decay into two
photons is furthermore sensitive to new charged, heavy particles of
physics beyond the SM. For these reasons a precise prediction of the
partial decay width $\Gamma(H \to \gamma \gamma)$ in the
intermediate-mass range is required. To this end the two-loop
calculations of the partial decay width have recently been completed.

Here, a short review of these  calculations is given focusing on the
EW contributions. Firstly, in section~2 the Born level results are
given. The individual EW contributions at two-loop order are discussed
in the following sections, namly the corrections due to light fermions
in section~3, the top-quark-induced corrections in section~4, and the
purely bosonic corrections in section~5. The sum of these
contributions together with the QCD corrections at two-loop order are
presented in section~6, which also contains the conclusion.


\section{Born level}
\begin{figure}[t]
\begin{center}
\epsfig{figure=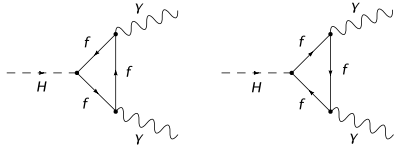,height=2.cm}
\end{center}
\caption{\label{haa1ltop} One-loop diagrams with virtual fermions.}
\end{figure}

\begin{figure}[t]
\begin{center}
\epsfig{figure=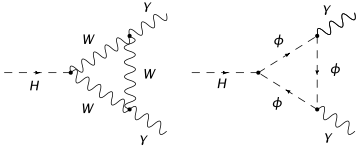,height=2.cm}
\end{center}
\caption{\label{haa1lW} One-loop sample diagrams with virtual bosons.}
\end{figure}

At Born level there exist two diagrams for each
fermion (Fig.~\ref{haa1ltop}). However, only the heavy top-quark contributes
sizeably while the other contributions are negligible. In addition one
has to consider 26 diagrams with virtual W bosons, Goldstone bosons,
and ghosts (Fig.~\ref{haa1lW}). Since heavy particles give sizeable
contributions to this loop-induced process, it is sensitive to new
charged particles of physics beyond the SM.

The exact results for the decay rate have been known since
long~\cite{Ellis:1975ap}. However, it is instructive to have a look at
the expansion of these results in the external momenta. Expansions
will partly be performed also at the two-loop level, where the exact
results are not known. A natural choice for the expansion parameters
is $\tau_t=M_H^2/(4 M_t^2)$ and $\tau_W=M_H^2/(4 M_W^2)$, respectively. It
turns out that the approximation consisting of the first three terms
of the expansion in $\tau_{t}$ is practically indistinguishable from
the exact result up to $\tau_t\approx 0.25$ for the case of the
diagrams with virtual top-quarks. In the second case, the convergence
is slightly worse, since $M_H=2M_W$ corresponds to $\tau_W=1$ and the
exact result behaves like $\sqrt{1-\tau_{W}}$ in this
limit. Nevertheless, for $M_H=120$~GeV, 140~GeV, and $2 M_W$, the
approximation by five expansion terms deviates from the exact result
by as little as 0.3\%, 1.6\%, and 19.9\%, respectively.

In the intermediate-mass regime of the Higgs boson the contribution
from virtual bosons and ghosts dominates, while it is partially
cancelled by the contribution due to virtual top-quarks.


\section{Corrections due to light fermions}
\begin{figure}[t]
\begin{center}
\epsfig{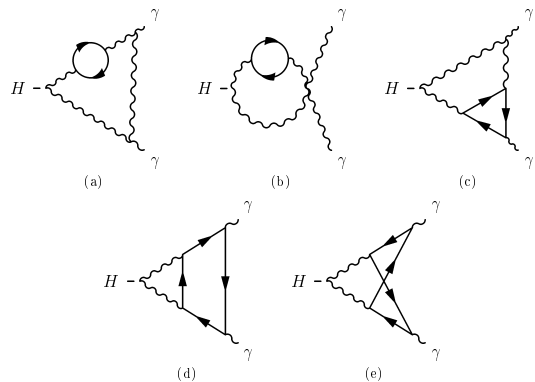}
\end{center}
\caption{\label{LFNLOdias} Generic two-loop diagrams for corrections
  due to light fermions; taken from~\cite{Aglietti:2004nj}.}
\end{figure}

\begin{figure}[t]
\begin{center}
\epsfig{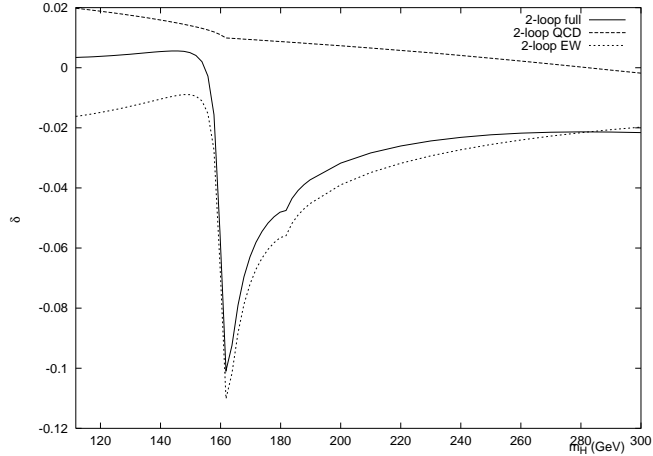}
\end{center}
\caption{\label{LFges} Relative EW corrections due to light fermions
  and QCD corrections as functions of the Higgs boson mass; taken
  from~\cite{Aglietti:2004nj}.}
\end{figure}

Light fermions can contribute at the two-loop level since it is
possible to avoid the direct coupling of the fermions to the Higgs
boson. The relevant diagrams are shown in
Fig.~\ref{LFNLOdias}. Furthermore, one has to sum over all generations
of light fermions. Therefore one can expect a non-negligible
contribution due to light fermions. The respective calculation was
done in~\cite{Aglietti:2004nj}. In this case an Asymptotic Expansion
is not possible since thresholds occur at $M_H=M_W$ and at $M_H=2 M_W$,
while we consider a Higgs boson mass just in this mass range. For this
reason an analytic calculation was performed. The authors considered
the limit of vanishing masses for the light fermions and employed the
Background Field Method (BFM) quantisation framework in order to reduce
the number of diagrams. Then they projected out the scalar amplitudes
and reduced them to a set of linearly independent ones. These were in
turn reduced to a set of master integrals by means of the
Integration-By-Parts technique. Finally the master integrals were
calculated using differential equations and the results were expressed
in terms of Generalised Harmonic Polylogarithms. The unphysical
singularity at the 2W-threshold was regularised through the
introduction of the width of the W boson by performing the replacement
$M_W \to M_W - i \Gamma_W/2$. It could be shown that the result is
independent on the regulator except in the region between 150~GeV and
170~GeV, where the result has to be taken with some caution. The
relative correction as a function of the Higgs boson mass is shown in
Fig.~\ref{LFges}. For comparison also the two-loop QCD result is shown
and the sum of these corrections. It turns out that in the
intermediate-mass region of the Higgs boson the corrections due to
light fermions are small but indeed non-negligible, between 1\% and~2\%.

\section{Top-quark-induced corrections}
\begin{figure}[t]
\begin{center}
\epsfig{figure=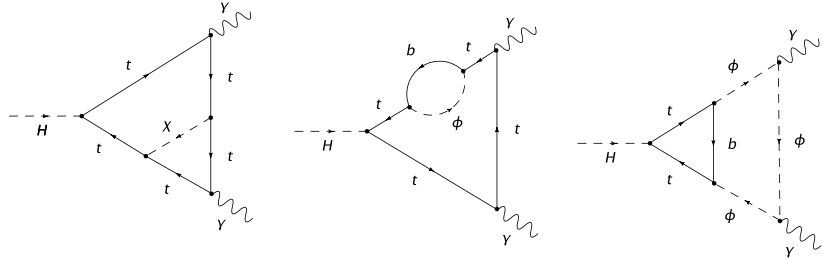,height=3.5cm}
\end{center}
\caption{\label{haa2ltop} Two-loop sample diagrams for top-quark-induced corrections.}
\end{figure}

\begin{figure}[hb]
\begin{center}
\epsfig{figure=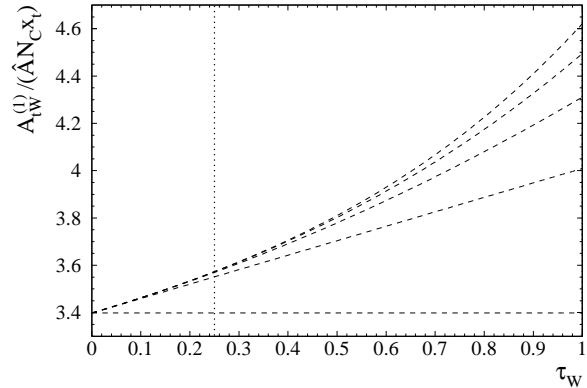,height=6cm}
\end{center}
\caption{\label{Topges} Normalised amplitude for the top-quark-induced
  two-loop electroweak corrections proportional to $G_F M_t^2$ as a
  function of $\tau_W$. The dashed curves represent the
  sequence of approximations that are obtained by successively
  including higher powers of $\tau_{W}$ in the expansion. The dotted
  vertical line and the right edge of the frame encompass the
  intermediate-mass range of the Higgs boson; taken
  from~\cite{Fugel:2004ug}.}
\end{figure}

\begin{figure}[hb]
\begin{center}
\epsfig{figure=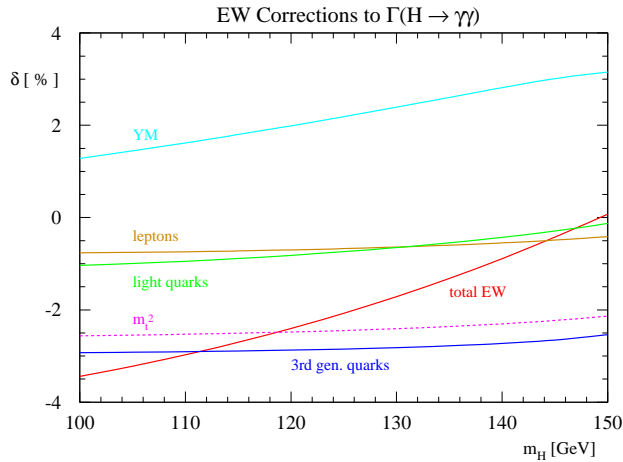,height=10cm,angle=270}
\end{center}
\caption{\label{NLOEWges} Individual relative EW corrections as
  functions of the Higgs boson mass; taken
  from~\cite{Degrassi:2005mc}.}
\end{figure}

The top-quark-induced corrections have first been considered in the
limit of a large top-quark mass in~\cite{Fugel:2004ug}. Sample
diagrams for this class of corrections are depicted in
Fig.~\ref{haa2ltop}. In order to obtain the correction of order
$\mathcal{O}$($G_F m_t^2$) as an expansion in $\tau_W$ up to and
including the terms of order $\mathcal{O}$($\tau_W^4$) the
Asymptotic Expansion technique was applied taking the bottom-quark to
be massless. Furthermore, the on-shell-scheme, dimensional
regularisation, the anticommuting definition of $\gamma_5$, and a
general $R_\xi\,$-gauge have been employed. Also the Tadpole diagrams
had to be included since the $m_t^4$-terms cancel out non-trivially in
the sum of contributions from genuine Tadpole diagrams, counterterms
and non-trivial terms in the Asymptotic Expansion. In
Fig.~\ref{Topges} the normalised amplitude is shown as a function of
$\tau_W$. It turns out that the convergence behaviour is similar to
the one at Born level for the diagrams with virtual bosons and
ghosts. For this reason it was assessed that the approximation should
be very good for Higgs boson masses up to 140~GeV and still reasonably
good up to the right edge of the intermediate-mass regime. By now also
the full top-quark-induced corrections have been completed
in~\cite{Degrassi:2005mc}, where also the leading term in the top-quark
mass could be recovered. The method of the calculation is the same as
in the case of the purely bosonic corrections and will be reviewed in
the respective section. In Fig.~\ref{NLOEWges} the result is shown as
the relative correction to the Born result as a function of the Higgs
boson mass. The correction lies in the range between 2.5\%
and~3\%. Also the result for the leading term is shown in this
figure. It is obviously a very good approximation to the full result.

\section{Purely bosonic corrections}
\begin{figure}[t]
\begin{center}
\epsfig{figure=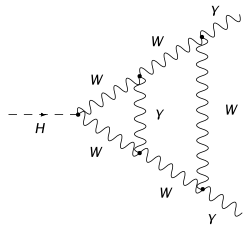,height=3cm}
\end{center}
\caption{\label{haa2lbos} Two-loop sample diagram for purely bosonic corrections.}
\end{figure}

Finally the purely bosonic corrections have to be taken into
account, one sample diagram of which is shown in
Fig.~\ref{haa2lbos}. The calculation of these corrections was also
performed in~\cite{Degrassi:2005mc}. The authors employed the BFM
quantisation framework and used the Tadpole counterterm in order to
cancel the diagrams containing a Tadpole. They projected out the
relevant form factor and performed a Taylor expansion in the parameter
$q_W=q^2/(4 M_W^2)$, where $q$ is the external momentum of the
Higgs boson, up to and including terms of order
$\mathcal{O}$($q_W^3$). The gauge parameter was renormalised in order
to obtain finite terms in this expansion which in turn allows for an
improvement of the results by means of a Pad\'e approximation. The
respective result is also shown in Fig.~\ref{NLOEWges}, where it is
denoted as YM. Its value is around 2\% with a sign opposite to the
corrections discussed above.

\section{Resulting NLO corrections and conclusion}
\begin{figure}[ht]
\begin{center}
\epsfig{figure=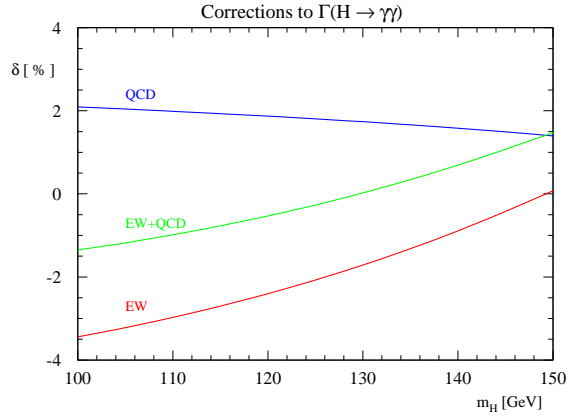,height=9cm,angle=270}
\end{center}
\caption{\label{NLOges} Full relative EW and relative QCD corrections
  as functions of the Higgs boson mass; taken
  from~\cite{Degrassi:2005mc}.}
\end{figure}

The purely bosonic corrections have a different sign compared to the
other corrections as is the case at Born level. Therefore a
partial cancellation takes place between the individual EW
contributions as can also be seen in Fig.~\ref{NLOEWges}. The QCD
corrections at the two-loop level (see~\cite{Zheng:1990qa}, for the
result at three loops see~\cite{Steinhauser:1996wy}) are also small
and cancel partially against the resulting EW corrections. This is
shown in Fig.~\ref{NLOges}. The reason for the smallness of the QCD
result could be due to the fact that only the Born level diagrams
containing virtual quarks are affected by QCD corrections. Note, that
also in the case of the QCD corrections at two loops a naive expansion
in the external momenta is possible, again leading to a rapidly
converging series as is the case for the Born level contributions
of the diagrams with virtual top-quarks.
The sum of the complete EW corrections ranges from $\sim -4\%$ to $\sim
0\%$ for a Higgs boson mass between 100~GeV and 150~GeV. The full
two-loop result amounts to less than $\pm 1.5\%$. The NLO calculation
therefore already gives a very reliable prediction.



\end{document}